\documentclass[12pt,journal,draftcls,letterpaper,onecolumn]{article}
\usepackage{wrapfig}
\usepackage{epsf}
\usepackage{graphicx}
\usepackage{subfigure}

\begin{document}

\title{Integration of Differential Gene-combination Search and Gene Set Enrichment Analysis: A General Approach}
\author{Gang Fang$^{1}\footnote{Corresponding author; Supplementary material: http://vk.cs.umn.edu/ICG/}$, Michael Steinbach$^{1}$, \\
Chad L. Myers$^{1}$, Vipin Kumar$^{1}$\\
$^{1}$Department of Computer Science, \\
University of Minnesota, Minneapolis, MN 55455, USA.}

\date{}

\maketitle

\begin{abstract}

Gene Set Enrichment Analysis (GSEA) and its variations aim to discover collections of genes that show moderate but coordinated differences in expression. However, such techniques may be ineffective if many individual genes in a phenotype-related gene set have weak discriminative power. A potential solution is to search for combinations of genes that are highly differentiating even when individual genes are not. Although such techniques have been developed, these approaches have not been used with GSEA to any significant degree because of the large number of potential gene combinations and the heterogeneity of measures that assess the differentiation provided by gene groups of different sizes.

To integrate the search for differentiating gene combinations and GSEA, we propose a general framework with two key components: (A) a procedure that reduces the number of scores to be handled by GSEA to the number of genes by summarizing the scores of the gene combinations involving a particular gene in a single score, and (B) a procedure to integrate the heterogeneous scores from combinations of different sizes and from different gene combination measures by mapping the scores to p-values. Experiments on four gene expression data sets demonstrate that the integration of GSEA and gene combination search can enhance the power of traditional GSEA by discovering gene sets that include genes with weak individual differentiation but strong joint discriminative power. Also, gene sets discovered by the integrative framework share several common biological processes and improve the consistency of the results among three lung cancer data sets.


\end{abstract}

\section{Introduction}
\label{sec:intro}

Microarray technology is an important tool to monitor gene-expression in bio-medical studies \cite{schena1995quantitative}. A common experimental design is to compare two sets of samples with different phenotypes, e.g. diseased and normal tissue, with the goal of discovering differentially expressed genes \cite{golub1999allaml}. Statistical testing procedures, such as such as the t-test and significance analysis of microarrays \cite{tusher2001significance}, have been extensively studied and widely used. Subsequently, multiple testing corrections are usually applied \cite{dudoit2003multiple}. A comprehensive review of such approaches are presented in \cite{pan2002comparative}.

Differential expression analysis based on univariate statistical tests has several well-known limitations. First, due to the low sample size, high dimensionality and the noisy nature of microarray data, individual genes may not meet the threshold for statistical significance after a correction for multiple hypotheses testing \cite{subramanian2005gsa}. Second, the lists of differentially expressed genes discovered from different studies on the same phenotype have little overlap \cite{subramanian2005gsa}. %

\begin{figure}[t]%
\centering
\includegraphics[width=.7\textwidth]{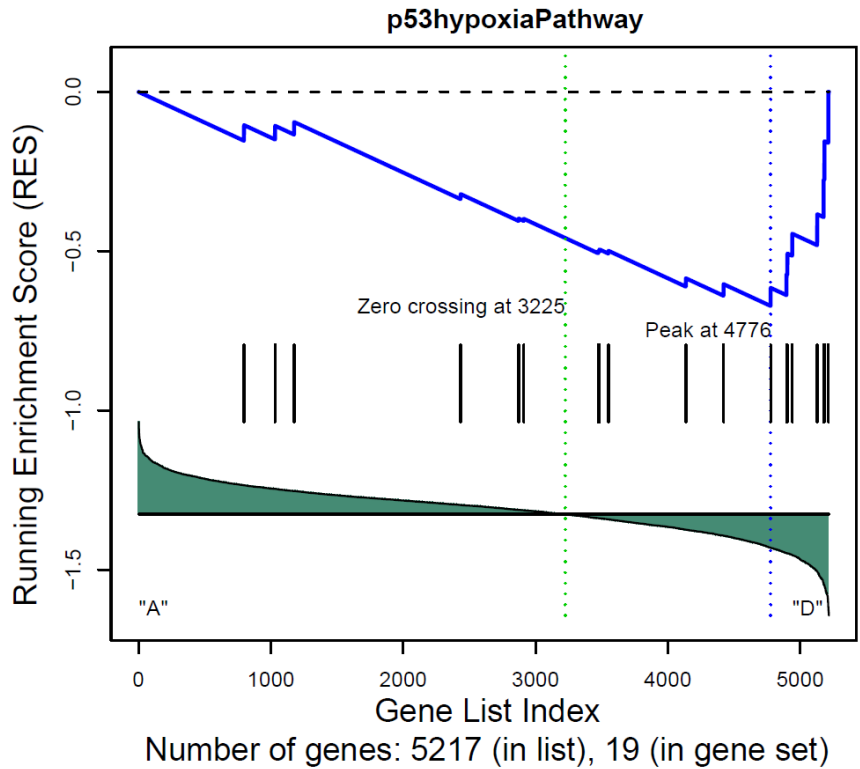}
\caption{A gene set (p53 hypoxial pathway) with many moderate but coordinated differential expression (towards the right tail of the ranked list). Figure generated with GSEA software \protect\cite{subramanian2005gsa}}
\label{fig:pnas2005toy}
\end{figure}

These limitations motivated the creation of Gene Set Enrichment Analysis GSEA \cite{mootha2003pgc,subramanian2005gsa}, which discovers collections of genes, for example, known biological pathways \cite{subramanian2005gsa}, that show moderate but coordinated differentiation. For example, Subramanian and Tamayo et al. \cite{subramanian2005gsa} report that the p53 hypoxial pathway contains many genes that show moderate differentiation between two lung cancer sample groups with different phenotypes. Although the genes in the pathway are not individually significant after multiple hypothesis correction \cite{subramanian2005gsa}, the pathway is. For those familiar with GSEA and its output, Figure \ref{fig:pnas2005toy} shows the GSEA results for the p53 hypoxial pathway. GSEA also has the advantages of better interpretability and better consistency between the results obtained by different studies on the same phenotype \cite{subramanian2005gsa}. Ackermann and Strimmer presented a comprehensive review of different GSEA variations in \cite{ackermann2009gmf261}.

\begin{figure}[t]
\centering
\subfigure[\small $M_1$\label{fig:gb2005toy1}]{\includegraphics[width=.48\textwidth]{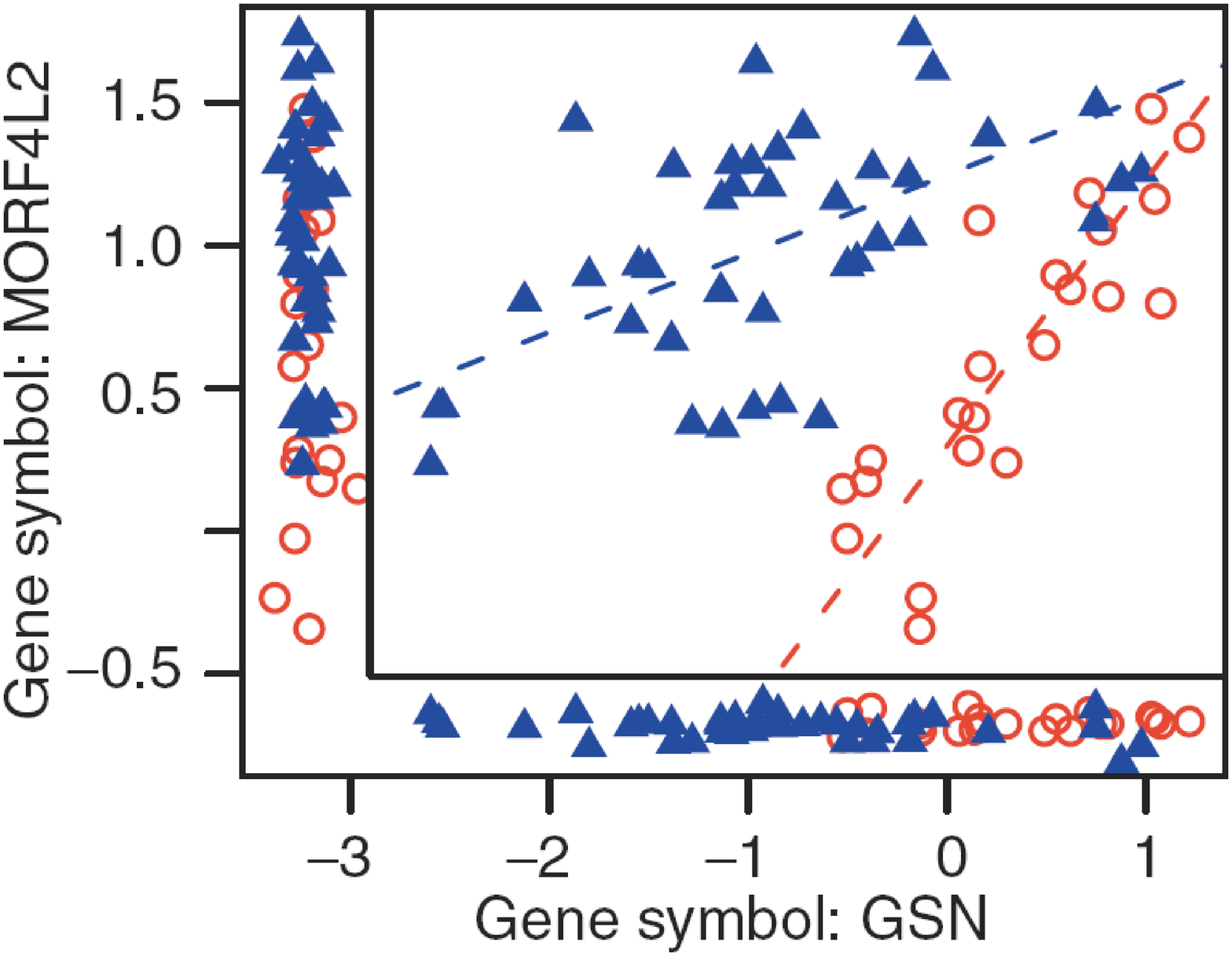}}
\subfigure[\small $M_2$\label{fig:gb2005toy2}]{\includegraphics[width=.48\textwidth]{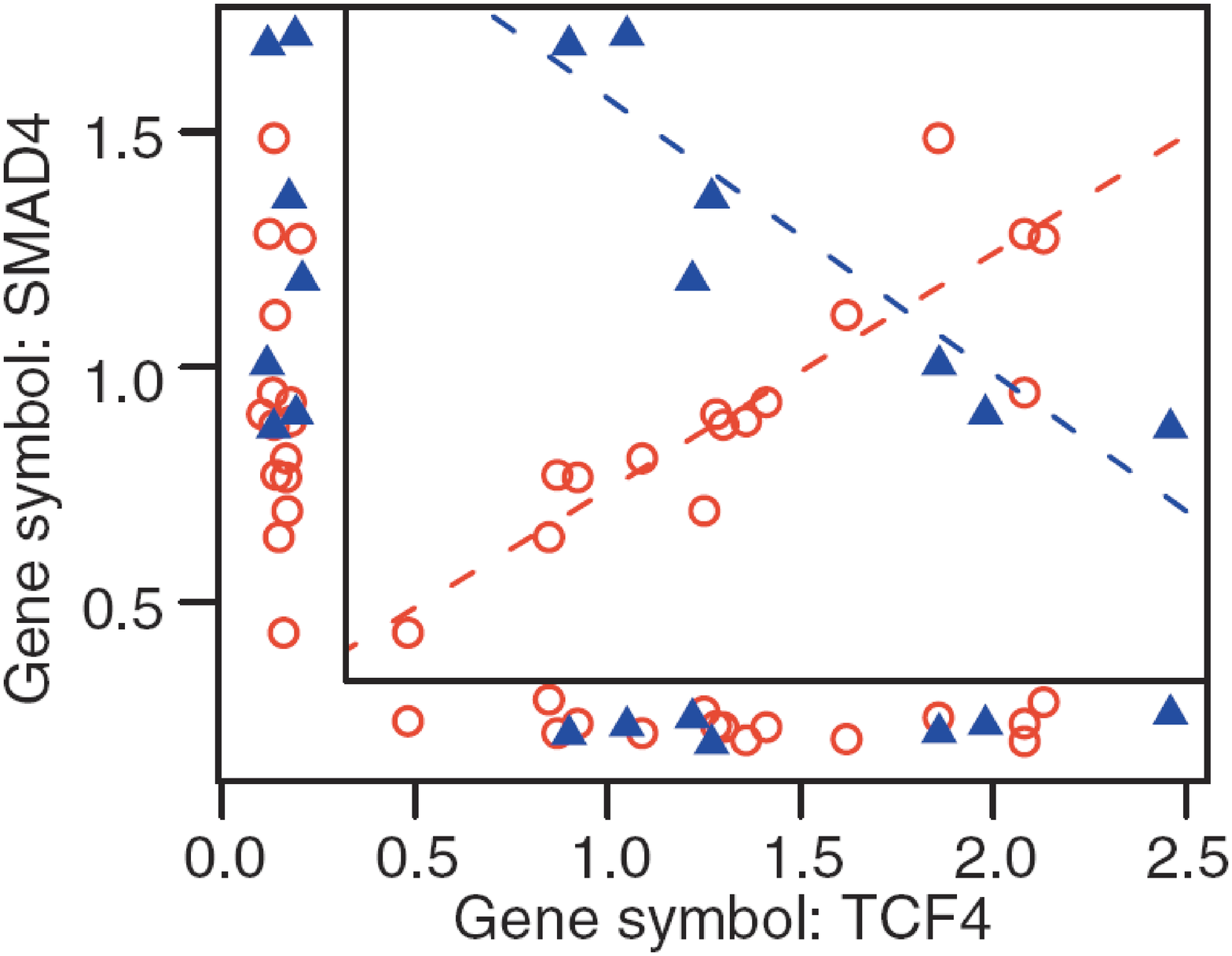}}
\caption{\small Two highly differential gene-pairs with weak individual discriminative power. Axes indicate the expression level of indicated genes. Different color and shape of markers indicates the two phenotypes. Figures modified from Dettling et al. \protect\cite{dettling2005searching}. This two types of differential gene combinations are measured respectively by two measures $M_1$ and $M_2$ as described in section \ref{sec:measures}.}
\label{fig:gb2005toy12}
\end{figure}

Unfortunately, GSEA and related techniques may be ineffective if many individual genes in a phenotype-related gene set have weak discriminative power. A potential solution to this problem is to search for combinations of genes that are highly differentiating even when individual genes are not. For this approach, the targets are groups of genes that show much stronger discriminative power when combined together \cite{dettling2005searching}. For example, Figure \ref{fig:gb2005toy1} illustrates one type of differentially expressed gene combination discovered in \cite{dettling2005searching}. The two genes have weak individual differentiation indicated by the overlapping class symbols on both the two axes. In contrast, these two genes are highly discriminative in a joint manner indicated by the different correlation structure in the two-dimensional plot, i.e. they are correlated along the blue and red dashed line respectively in the triangle and circle class. Such a joint differentiation may indicate that the interaction of the two genes is associated with the phenotypes even though the two genes, individually, are not.

Figure \ref{fig:gb2005toy2} illustrates another type of phenotype-associated gene combination discovered in \cite{dettling2005searching}, usually named differential coexpression \cite{kostka2004fds,fang2010sdc}, in which the correlation of the two genes are high in one class but much lower in the other class. As discussed in \cite{dettling2005searching,glazko2009unite}, existing multivariate tests such as Hotelling's $T^2$ \cite{lu2005hotelling}, Dempster's T1 \cite{dempster1958high} are not suitable to detect such `complementary' gene combinations because they only screen for differences in the multivariate mean vectors, and thus will favor pairs that consist of genes with strong marginal effects by themselves but not the genes like the four in Fig. \ref{fig:gb2005toy12}. For clarification, we use differential gene combination search (denoted as DGCS) to refer to the multivariate data analyses that are designed to detect the complementarity of different genes, rather than those designed to model the correlation structure of different genes (such as Hotelling's $T^2$ and Dempster's T1 test). A variety of other DGCS measures for complementary gene combination search are proposed for gene pairs \cite{li2002gwc,lai2004smi} in addition to the two illustrated in figure \ref{fig:gb2005toy12}. Several measures are designed for higher-order gene combination beyond pairs \cite{fang2010sdc,klebanov2007nstat}. These approaches can provide biological insights beyond univariate gene analysis as shown in \cite{dettling2005searching,fang2010sdc,klebanov2007nstat}.

The limitations of GSEA and the capabilities of DGCS motivate a GSEA approach using gene combinations in which the score of a gene set is based on both the scores of individual genes in the set and the scores from the gene combinations in which these genes participate. Unfortunately, gene combination techniques have not been used with the GSEA approach in any significant way because of two key challenges.
\begin{enumerate}
  \item \textbf{Finding a technique to reduce the vast number of gene combinations.} There are exponentially more gene combinations than individual genes, i.e. in addition to the $N$ univariate genes, there are $N^2$ gene-pairs, $N^3$ gene-triplets, etc. Many variations of GSEA are based on a ranked list of the $N$ individual genes as illustrated in Figure \ref{fig:pnas2005toy}. Including combinations in the ranked list might work for size-2 combinations \cite{choi2009statistical,zhang2009identifying}, but would not be feasible for handling gene combinations of larger sizes. Furthermore, this explosion in the number of gene combinations negatively impacts false discovery rates. Thus, by adding so many gene combinations, we run the risk that neither groups of genes nor individual genes will show statistically significant differentiation.

  \item \textbf{Combining results from the heterogeneous measures used to score different size gene combinations.} Furthermore, because a gene can be associated with the phenotype either as an univariate variable or together with other genes as a combination, the importance of a gene set should be based on both the univariate gene scores and the gene combination based scores of its set members. However, different measures have a different nature, scale and significance, and thus are not directly comparable (to be detailed in section \ref{sec:score2pv}). Indeed, differences exist even between gene combinations of the same measure but of different sizes. Therefore, the challenge lies in how to design a framework to combine different measures (a univariate measure\footnote{Ackermann and Strimmer \cite{ackermann2009gmf261} suggest that different univariate statistics have similar effect in GSEA and thus, we consider only one univariate statistic rather than multiple of them.} plus one or more DGCS measures) together within the GSEA framework.

\end{enumerate}

To the best of our knowledge, no existing work has sufficiently addressed these two challenges, although recent work presented in \cite{choi2009statistical,zhang2009identifying} have made initial efforts at adding GSEA capabilities to gene combination search. More specifically, two approaches are proposed in \cite{choi2009statistical,zhang2009identifying} to help the study of a specific type of size-2 differential combinations as illustrated in \ref{fig:gb2005toy2}. The experiments in these two studies provide some evidence about the benefits of the integration. However, a more general framework is needed that can also handle other types of size-2 differential combinations as illustrated in figure \ref{fig:gb2005toy1}, higher order differential combinations (e.g. SDC\cite{fang2010sdc} and the n-statistic \cite{klebanov2007nstat}), and multiple types of differential combinations.

\textbf{Contributions:} In this paper, we propose a general framework to address the above challenges for the effective integration of DGCS and GSEA. Specific contributions are as follows:
\begin{enumerate}
	
\item \textbf{A gene-combination-to-gene score summarization procedure (procedure \emph{A}) that is designed to handle the exponentially increasing number of gene combinations.} First, for a given gene combination measure and a certain $k$, the score of a size-$k$ combination is partitioned into $k$ equal parts which are assigned to each of the $k$ genes in the combination. Because each gene can participate in up to ${N-1 \choose k-1}$ size-$k$ combinations, each gene will be assigned with a score from each of these combinations. Secondly, an aggregation statistic, e.g., maximum absolute value is used to summarize the different scores for a gene. With such a procedure, scores for all the size-$k$ gene combinations are summarized to $N$ scores for $N$ genes. This procedure can effectively retain the $O(N)$ length of the ranked list while handling gene combinations of size-$k$ ($k \geq 2$).
	
\item \textbf{A score-to-pvalue transformation and summarization procedure (procedure \emph{B}) that is designed to integrate the scores contributed (in procedure $A$) from different gene combination measures and from gene combinations of different sizes.} The transformation is based on p-values obtained from scores derived from phenotype permutations. Such a transformation enables the comparison of scores from different measures (either univariate or gene combination measures) and scores from the gene combinations of different sizes. Subsequently, among all the p-values of a gene, the best is used as an integrated score of statistical significance.

\item \textbf{Integration of the above two procedures with GSEA} More specifically, after procedures \emph{A} and \emph{B}, each gene has a single integrated score. Unlike traditional univariate scores, these $N$ integrated scores are based on both the univariate statistic and the gene combination measures. For the type of GSEA variations that depend on phenotype permutation test, $P+1$ lists of $N$ integrative scores are computed, one for the real class labels and the other for the $P$ permutations. For the type of GSEA variations that are based on gene-set permutation test, only the list of integrated scores for the real class labels are needed. An independent Matlab implementation of the proposed framework is available for download, which allows most existing GSEA frameworks \cite{ackermann2009gmf261} to directly utilize the proposed framework to handle gene combinations with almost-zero modification.  %

\item \textbf{Experimental results that illustrate the effectiveness of the proposed framework.} We integrated three gene combination measures and the GSEA approach presented in \cite{subramanian2005gsa} and produced experimental results from four gene expression datasets. These results demonstrate that the integrative framework can discover gene sets that would have been missed without the consideration of gene combinations. This includes statistically significant gene sets with moderate differential gene combinations whose individual genes have very weak discriminative power. Thus, a gene combination assisted GSEA approach can improve traditional GSEA approaches by discovering additional disease-associated gene sets. Indeed, the integrative approach also improve traditional DGCS since most gene combinations are not statistically significant by themselves. Furthermore, we also show that the biologically relevant gene sets discovered by the integrative framework share several common biological processes and improve the consistency of the results among the three lung cancer data sets.

\end{enumerate}

\textbf{Overview:} The rest of the paper is organized as follow. In section \ref{sec:measures}, we describe three gene combination measures used in the following discussion and experiments. In Section \ref{sec:methods}, we present the technical details of the two procedures of the general integrative framework. Experimental design and results are presented in Section \ref{sec:exp}, followed by conclusions and discussions in Section \ref{sec:discssion}.

\section{Differential Gene Combination Measures}
\label{sec:measures}

In this section, we describe three DGCS measures for use in the following discussion and experiments. Let $A = \left\{a_1,a_2,\ldots,a_{|A|} \right\}$ and $B = \left\{b_1,b_2,\ldots,b_{|B|} \right\}$ be two phenotypic classes of samples of size $|A|$ and $|B|$ respectively. For each sample in $A$ and $B$, we have the expression value of $N$ genes $\textbf{G} = \left\{G_1,G_2,\ldots,G_N \right\}$. First, we have the following two measures (denoted as $M_1$ and $M_2$) defined for a pair of genes as presented in \cite{dettling2005searching}:

\begin{equation}
\scriptsize
M_1(G_i,G_j) = corr_{A}(G_i,G_j) + corr_{B}(G_i,G_j) - corr_{A \cup B}(G_i,G_j)
\label{eq:m1}
\end{equation}

\begin{equation}
\scriptsize
M_2(G_i,G_j) = corr_{A}(G_i,G_j) - corr{B}(G_i,G_j)
\label{eq:m2}
\end{equation}

where $G_i$ and $G_j$ are two genes, and $corr_D(G_i,G_j)$ represents the correlation of $G_i$ and $G_j$ over the samples in set $D$. As discussed in \cite{dettling2005searching}, $M_1$ and $M_2$ can detect the joint differential expression of two genes as illustrated in Figure \ref{fig:gb2005toy1} and Figure \ref{fig:gb2005toy2} respectively. $M_1$ and $M_2$ are used as two representative measures for gene pairs. Other options for gene combination measures for gene pairs have been investigated in \cite{lai2004smi,li2002gwc}.

We use the subspace differential coexpression measure (denoted as $M_3$) proposed in \cite{fang2010sdc} as the representative for measures for size-$k$ gene combinations, where $k$ can be any integer ($k \geq 2$).

\begin{equation}
\scriptsize
M_3(\alpha) = R_A(\alpha) - R_B(\alpha)
\label{eq:m3}
\end{equation}

where $\alpha$ is a set of genes such that $\alpha \subseteq G$ and $|\alpha| = k$. $R_A(\alpha)$ and $R_B(\alpha)$ respectively represent the fraction of samples in $A$ and $B$ over which the genes in $\alpha$ are coexpressed. $M_3$ is a generalization of $M_2$ for detecting the differential coexpression of $k$ genes ($k \geq 2$), i.e. the $k$ genes are highly coexpressed over many samples in one class but over far fewer samples in the other. Other options for size-k combinations include the \emph{n-statistic} \cite{klebanov2007nstat}, \emph{SupMaxPair} \cite{smk200904tech}, etc.

Signal-to-noise ratio (denoted as $M_0$) is used as the representative of traditional univariate statistics as in \cite{subramanian2005gsa,mootha2003pgc}.

\begin{equation}
\scriptsize
M_0(G_i) = \frac{\mu_A(G_i)-\mu_B(G_i)}{\sigma_A(G_i)+\sigma_B(G_i)}
\label{eq:m4}
\end{equation}

where $\mu_A(G_i)$ and $\mu_B(G_i)$ are the mean expression of $G_i$ in class $A$ and $B$ respectively, and $\sigma_A(G_i)$ and $\sigma_B(G_i)$ are the standard deviation of the expression of $G_i$ in class $A$ and $B$ respectively. Many other univariate statistics can be found in \cite{ackermann2009gmf261}.

In this paper, these four measures are used as representatives of each category for the illustration of the proposed integrative framework. However, the framework is general enough to handle other measures from each of these categories.

\section{Methods}
\label{sec:methods}

In Section \ref{sec:intro}, we motivated the integration of DGCS with GSEA, discussed two challenges associated with this integration, and briefly described two main procedures in the proposed framework. In this section, we present the technical details of the two procedures and their integration with GSEA.

\subsection{Procedure $A$: combination-to-gene score reduction}
\label{sec:procA}

There are two steps in procedure $A$. In step $(1)$, for each DGCS measure and each size-$k$ gene combination, its score is divided into $k$ equal parts and assigned to each of the $k$ genes in the combination. In step $(2)$, the scores assigned to a gene from all the size-k combinations in which the gene participates are summarized into a single score by an aggregation functions such as $max$. Note that, for most univariate statistic and DGCS measures which can be either positive or negative (e.g. the four measures described in section \ref{sec:measures}), the maximum is taken over the absolute values of the scores, and the sign of the score with the highest absolute value is recorded for later use. Other simple statistics such as mean or median, or sophisticated ones such as weighted summation \cite{junzhu2009ploscbcoexpression} can also be used. Since the focus of this paper is the overall integrative framework, we use $max$ for simplicity.

We provide a conceptual example of procedure $A$ for a gene $G_1$ with a certain DGCS measure $M_1$. This example considers gene combinations up to size $4$ for illustration purpose. The gene is associated with scores assigned from gene combinations of size 2, 3 and 4 (denoted as $C_2$, $C_3$ and $C_4$ respectively) in which $G_1$ participates. In step $(2)$, the scores from $C_2$, $C_3$ and $C_4$ are summarized by three maximum values, respectively. Please refer to the Appendix section for the illustration of this example.%

Procedure $A$ serves as a general approach to summarize the ${N \choose k}$ scores of all the size-$k$ combinations into $N$ scores for the $N$ genes. If we want to integrate GSEA with one DGCS measure and a specific size-$k$, procedure $A$ by itself can enable most existing variations of GSEA to search, with almost-zero modification, for statistically significant gene sets with moderate but coordinated gene combinations of size-$k$. Such a GSEA approach can collectively consider the gene combinations affiliated with a gene set, and may provide better statistical power and better interpretability for DGCS, as will be shown in the experiments.

\subsection{Procedure $B$: Score-to-pvalue conversion and summarization}
\label{sec:score2pv}

The hypothesis tested when one DGCS measure, say $M_x$, is integrated with GSEA (by procedure $A$) is that, whether a gene set includes significantly many genes with highly positive (or highly negative) combination-based scores measured by $M_x$. An extended hypothesis can be whether a gene set includes significantly many genes with highly positive (or highly negative) scores, either univariate or combination-based scores measured by different DGCS measures. The biological motivation of this extended hypothesis is that, a gene can be associated with the phenotype either as an univariate variable or together with other genes as a combination. To test this extended hypothesis, we design a second procedure (\emph{B}) that can integrate the scores of a gene from different measures.

Before describing the steps in this procedure. We first discuss in detail the challenges of integrating heterogeneous scores from different DGCS measures and combinations of different sizes.

\begin{enumerate}

	\item \textbf{The different nature of different measures}: Different measures are designed to capture different aspects of the discriminative power of a gene or a gene combination between the two phenotypic classes. Signal-to-noise ratio ($M_0$), a univariate gene-level statistic, measures the difference between the means of the expression of a gene in the two classes. In contrast, $M_2$, a differential coexpression measure for a pair of genes describes the difference of the correlations of a gene-pair in the two classes. Thus, for a gene, the score of itself measured by $M_0$ and the score assigned and summarized from the ${N-1 \choose 1}$ size-2 gene combinations measured by $M_2$ are not directly comparable. Similarly, the scores of different DGCS measures can also have a different nature, e.g. $M_1$ and $M_2$ as illustrated in figure \ref{fig:gb2005toy12}.
	
	\item \textbf{The different scales of different measures}: Different measures also have different ranges of values. For example, the range of $M_0$, $M_1$, $M_2$, and $M_3$ are $\left[-\infty, \infty\right]$, $\left[-3, 3\right]$, $\left[-2, 2\right]$ and $\left[-1, 1\right]$ respectively. Thus, they are not directly comparable.
		
	\item \textbf{Differences in significance between different measures}: Even after we normalize the scores of different measures to a single range, say  $\left[-1, 1\right]$, they are still not comparable because the scores of different measures have different statistical significance. For example, a normalized $M_0$ score of $0.8$ may be less significant than a normalized $M_1$ score of 0.5, if there are many genes with normalized $M_0$ score greater than $0.8$ in the permutation test \cite{subramanian2005gsa}, but very few genes with normalized $M_1$ score greater than $0.5$ in the permutation test. Note that, such differences in statistical significance also exists between gene combinations of different sizes, even for the same measure. Take the subspace differential coexpression measure $M_3$ as an example. A score of $0.5$ for a size-2 combination may not be as significant as a score of $0.5$ for a size-3 combination as discussed in \cite{fang2010sdc}.
	
\end{enumerate}

To handle the above heterogeneity, we propose a score-to-pvalue transformation and summarization procedure that can enable the comparison and integration of the scores of different measures and combinations of different sizes. There are three major steps in procedure $B$.

\begin{figure}[t]
\centering
\includegraphics[width=0.95\textwidth]{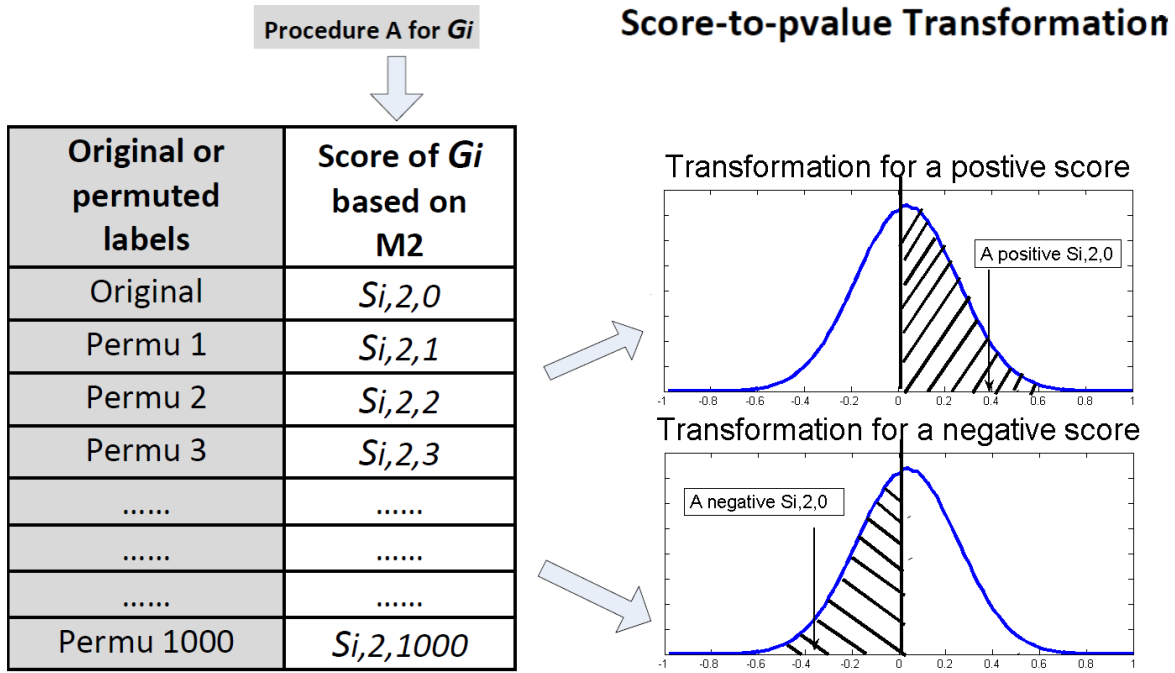}
\caption{Illustration of step $1$ in procedure $B$ (score-to-pvalue transformation) for gene $G_i$ and measure $M_2$.} %
\label{fig:tabmerge2measures}
\end{figure}

\subsubsection{\textbf{Step $1$: Score-to-pvalue transformation}}

Consider a concrete example. For a gene $G_i$ and a measure $M_2$, procedure $A$ computes a single summarized score. In this step, the original phenotype class labels are permutated say $1000$ times, and for each permutation, the same procedure $A$ is applied, and a corresponding score for $G_i$ and $M_2$ is computed. We denote the score of $G_i$ and $M_2$ summarized with the original label as $S_{i,2,0}$, where $i$ is the gene index, and $2$ indicates the measure and $0$ means it is the score based on the original label. Similarly, we denote the scores computed in each of the permutation as $S_{i,2,j}$, where $1 \leq j \leq 1000$.

These $1001$ scores are organized in the table on the left in figure \ref{fig:tabmerge2measures}. The $1000$ scores computed in the $1000$ permutations can be considered as the null-distribution for gene $G_i$ and measure $M_2$, and a p-value can be estimated for $S_{i,2,0}$. Specifically, if $S_{i,2,0}$ is positive, the p-value is the ratio of the number of scores that are greater or equal to $S_{i,2,0}$ and the number of scores that are positive. Similarly if $S_{i,2,0}$ is negative, the p-value is computed as the ratio of the number of scores which are less or equal to $S_{i,2,0}$ and the number of scores which are negative.\footnote{Treating positive and negative scores separately follows the practice of GSEA \cite{tian2005q1q2gsea}} Note that, such a score-to-pvalue transformation is done for both $S_{i,2,0}$ and each of $S_{i,2,j}$ ($1 \leq j \leq 1000$), if the GSEA approach to be integrated is based on phenotype permutation test \cite{tian2005q1q2gsea}. Otherwise, only $S_{i,2,0}$ needs to be transformed to p-value and will be used by the GSEA approaches that are based on gene-set permutation \cite{tian2005q1q2gsea}. In this paper, we illustrate the proposed framework using the GSEA approach presented by Subramanian and Tamayo et al. \cite{subramanian2005gsa} which is based on phenotype permutation test.

Essentially, step $1$ transforms the heterogeneous scores of a gene measured by different measures into their corresponding significance values, which are comparable to each other although their original values are not.

\subsubsection{\textbf{Step $2$: P-value Summarization}} Suppose that there are $Q$ different measures to be integrated, one of which is a univariate statistic, and the others are different DGCS measures for which we consider combinations of sizes up to $K$. After step $1$, each gene has a p-value for the univariate measure and up to $K$ p-values for each size of gene combination for each measure. In step $2$, the best\footnote{"Best" means it is the lowest raw p-value or the highest $-log_{10}$ transformed p-value} p-value associated with a gene is selected as the integrated significance.

Essentially, procedure $B$ integrates the scores of different DGCS measures for a gene and the univariate statistic of the gene into a single p-value. Such a statistical significance-based integration of heterogeneous scores enables the comparison and thus the ranking of all the $N$ genes. However, this ranked list does not maintain the original directionality of the integrated scores of each gene. In particular, most univariate statistics and DGCS measures (e.g. all the four measures described in section \ref{sec:measures}) can be either positive or negative. Such directionality information is lost in step $1$ and $2$ because the p-value is non-negative. Next, we describe a third step to maintain the directionality in the integration.

\subsubsection{\textbf{Step $3$: Maintaining directionality associated with the integrated p-values}}

In the simple case, the measures to be integrated capture the same type of differentiation between the two phenotype classes, e.g. $M_2$ and $M_3$. Suppose there are two genes $G_i$ and $G_j$, whose integrated p-values are transformed respectively from two scores measured by $M_2$ and $M_3$ in step $2$. The signs of these two scores are comparable to each other, because both $M_2$ and $M_3$ capture the change of coexpression of a combination of genes. Thus, we simply use the signs of these two scores as the signs associated with the integrated p-values of $G_i$ and $G_j$. Similarly, we associate a sign to all the $N$ integrated p-values. And these $N$ p-values with associated signs can be used to rank the $N$ genes based on their significance as well as their direction of differentiation, i.e. p-values associated with positive signs are ranked with descending significance, and afterwards, p-values associated with negative signs are ranked with increasing significance.

In the other case, if the measures to be integrated capture different types of differentiation between the two phenotype classes, the directionality can not be fully maintained. For example, suppose there are two genes $G_i$ and $G_j$, whose integrated p-values are transformed respectively from two scores measured by $M_0$ and $M_2$ in step $2$. The signs of these two scores are not comparable, because $M_0$ captures the change of mean expression, and $M_2$ captures the change of coexpression of a combination of genes. Specifically, up-regulation of $G_i$ can be associated with either high or low coexpression of another gene-combination in which $G_j$ participates. Thus, it is not reasonable to follow the same strategy to associate signs to the $N$ integrated p-values. If we know the correspondence of the signs of different genes in advance, e.g. the up-regulation of $G_1$ is associated with the low coexpression of $G_2$ and $G_3$, then the signs can be maintained. However, because it is not realistic to assume such prior knowledge, we propose the following heuristic approach which has proved a workable solution for our initial experiments. Specifically, since the focus of step $B$ is to integrate different DGCS measures in addition to the univariate statistic $M_0$, we considered $M_0$ as the base measure. For the integrated p-values that are transformed from scores measured by $M_0$ in step $2$ (say there are $w$ of them), we use the signs of these $w$ $M_0$ scores for the $w$ integrated p-values. For the signs of the other $N-w$ genes, we assign positive signs to all of them once and negative signs to all of them a second time. Correspondingly, we have two ranked lists similar to the simple case described above.

Note that, if the directionality of differential measures can be preserved, the power of this approach will be enhanced. To deal with the situation where signs are not comparable, other approaches will be explored.

\subsection{Integration with GSEA}

From the above description of procedure $A$ and $B$, we know that, if only one DGCS measure is used in the GSEA framework, only procedure $A$ is needed. If one or multiple DGCS measures are integrated together with the univariate statistic $M_0$ in the GSEA framework, procedure $B$ is needed in addition. In the first case, the integrative framework outputs a ranked list of $N$ scores with associated signs for the original class label, and $1000$ lists corresponding to the $1000$ permutation tests. In the second case, we have two sets of $1001$ lists respectively for the two rounds of maintaining directionality in step $3$ in procedure $B$.

In either case, the $1001$ ranked lists along with the appropriate parameter settings and specification of gene sets can be used to run GSEA. The only modification to GSEA is the elimination of the initial GSEA step to generate the scores, simulated and actual, that measure the level of differentiation between genes across different phenotypes. The proposed integrative framework is implemented as a Matlab function (available at http://vk.cs.umn.edu/ICG/), independently from the GSEA framework to be integrated in this paper \cite{subramanian2005gsa}. As summarized by Ackermann and Strimmer \cite{ackermann2009gmf261}, hundreds of variations of GSEA are being used by different research groups. This independently implemented integrative framework can be easily applied to other variations of GSEA. %

\subsection{A further technical detail}

In our experiments, in order to have a fair comparison, we transform the $1001$ ranked lists into the exact sample distribution as the original lists corresponding to $M_0\oplus$GSEA. Specifically, we only use the ranking information in the $1001$ integrated ranked lists and map to them to the values in the original lists based on $M_0\oplus$GSEA. Essentially, the values in the ranked list passed to the GSEA framework are exactly the same among $M_0\oplus$GSEA, $M_{01}\oplus$GSEA, $M_{02}\oplus$GSEA, $M_{03}\oplus$GSEA and $M_{0123}\oplus$GSEA, while the only difference is that the $N$ genes have different ranks in the lists. Such a mapping ensures that the additionally discovered gene sets are because of the integration of gene-combinations in addition to univariate statistic, rather than simply the different value distributions in the $1001$ lists.

\section{Results}
\label{sec:exp}

In this section, we present the experimental design and results for the evaluation of the proposed integrative framework. We first provide a brief description of the data sets and parameters used in the experiments. Second, we describe and discuss the comparative experiments to study whether the integration of DGCS and GSEA (denoted as DGCS$\oplus$GSEA) improves both DGCS and GSEA. The two major evaluation criteria are the statistical power to discover (additional) significant results, and the consistency of the results across different datasets for the study of the same phenotype classes. %

\subsection{Data sets}

The four datasets used in the experiments are described as follows:

\begin{enumerate}
	\item	Three lung cancer datasets respectively denoted as Boston~\cite{bhattacharjee2001lung}, Michigan~\cite{beer2002lung} and Standford~\cite{garber2001lung}: all the three data sets consist of gene-expression profiles in tumor samples from respectively 62, 86 and 24 patients with lung adenocarcinomas and provide clinical outcomes (classified as "good" or "poor" outcome). The two phenotypic classes in these three datasets are denoted as $A$ and $D$ as in \cite{subramanian2005gsa}. %

	\item A data set from the NCI-60 collection of cancer cell lines for the study of p53 status \cite{olivier2002iarc} (denoted as $P53$ data set): the mutational status of the p53 gene has been reported for 50 of the NCI-60 cell lines, with 17 being classified as normal and 33 as carrying mutations in the gene. The two phenotypic classes in this dataset are denoted as $MUT$ and $WT$ as in \cite{subramanian2005gsa}. %

\end{enumerate}

All four datasets were downloaded from the GSEA website\footnote{http://www.broadinstitute.org/gsea/}\cite{subramanian2005gsa}, and were already preprocessed as described in the supplementary file of \cite{subramanian2005gsa}. For all four data sets, we use the gene sets from $C_2$ in MSigDB$^4$ as in \cite{subramanian2005gsa}, as well as the same parameters. 


\subsection{Differential gene-combination measures}

We consider one univariate statistic ($M_0$), and three gene-combination measures ($M_1$, $M_2$ and $M_3$) in our experiments. These four measures are described in section \ref{sec:measures}. $M_1$ and $M_2$ are defined only for size-2 combinations. For $M_3$, we considered gene-combinations of size-$2$ and size-$3$ for the illustration of concept.

\subsection{Q1: Does GSEA-assisted DGCS improve traditional DGCS?}
\label{sec:integrative2DGCS}

\begin{table}[t]
\centering
\begin{tabular}{lllll}\hline
& Boston & Michigan & Stanford & P53\\\hline
$M_1$ & 0 & 2 & 0 & 0 \\
$M_2$ & 645 & 1 & 2 & 1 \\
$M_3$ & 10 & 1 & 0 & 0 \\\hline
\end{tabular}
\caption{Number of \underline{gene combinations} with FDR less than $0.25$ discovered from the four data sets by each combination measure}
\label{tab:DGCSFDRp25}
\end{table}

In this section, we study whether the question (Q1) of whether integration of DGCS and GSEA can improve traditional DGCS. For this comparison, we consider the integration of DGCS and GSEA as a GSEA-assisted DGCS approach. We first apply the traditional DGCS approaches on the four datasets to find statistically significant gene-combinations. We denote the three DGCS approaches respectively with the names of the three measures, i.e. $M_1$, $M_2$ and $M_3$. Second, we apply the integrative framework, in which GSEA is integrated respectively with the three DGCS measures, to find statistically significant gene sets with moderate but coordinated differential gene-combinations. We denote the three instances of the integrative approach respectively as $M_1\oplus$GSEA, $M_2\oplus$GSEA and $M_3\oplus$GSEA. Then, we compare the results of $M_1$, $M_2$ and $M_3$, respectively with the results of $M_1\oplus$GSEA, $M_2\oplus$GSEA and $M_3\oplus$GSEA.

\begin{table}[t]
\centering
\begin{tabular}{lllll}\hline
& Boston & Michigan & Stanford & P53\\\hline
$M_1\oplus$GSEA & 4$^{(A)}$ & 1$^{(A)}$ & 4$^{(A)}$ & 13$^{(PX)}$ \\
$M_2\oplus$GSEA & 1$^{(H)}$ & 7$^{(HS)}$ & 4$^{(HS)}$ & 0 \\
$M_3\oplus$GSEA & 0 & 1$^{(I)}$ & 3$^{(X)}$ & 2$^{(P)}$ \\\hline
\end{tabular}
\caption{Number of \underline{gene sets} with FDR less than $0.25$ discovered from the four datasets by integrating GSEA with each of the three combination measures. One or multiple biological process(es) are indicated as superscript, from which we can observe the consistency across three lung cancer data sets. $A$: apoptosis related pathways; $H$: responses to hypoxia; $S$: \emph{sppaPathway}; $I$: insulin-signaling sets; $X$: oxidative-phosphorylation related sets; $P$: p53-related sets. The names of the discovered gene sets and their FDRs are available in the Appendix section.}
\label{tab:gseaFDRp25}
\end{table}

In this section, we study whether the question (Q1) of whether integration of DGCS and GSEA can improve traditional DGCS. For this comparison, we consider the integration of DGCS and GSEA as a GSEA-assisted DGCS approach. We first apply the traditional DGCS approaches on the four datasets to find statistically significant gene-combinations. We denote the three DGCS approaches respectively with the names of the three measures, i.e. $M_1$, $M_2$ and $M_3$. Second, we apply the integrative framework, in which GSEA is integrated respectively with the three DGCS measures, to find statistically significant gene sets with moderate but coordinated differential gene-combinations. We denote the three instances of the integrative approach respectively as $M_1\oplus$GSEA, $M_2\oplus$GSEA and $M_3\oplus$GSEA. Then, we compare the results of $M_1$, $M_2$ and $M_3$, respectively with the results of $M_1\oplus$GSEA, $M_2\oplus$GSEA and $M_3\oplus$GSEA.

Table 1 lists the number of statistically significant gene combinations discovered respectively by the three measures on each of the four datasets, with an FDR threshold of $0.25$. Table 1 lists the number of statistically significant gene sets discovered by integrating GSEA respectively with the three DGCS measures on each of the four datasets, also with the same FDR threshold of $0.25$. Three major observations can be made by comparing the two tables:

\subsubsection{\textbf{GSEA-assisted DGCS has better statistical power than traditional DGCS}}

Table 1 shows that, in most cases, traditional DGCS discovers very few (less than 3) statistically significant gene combinations (although $M_2$ and $M_3$ have $645$ and $10$ gene-combinations on the Boston data set, none of them have FDR lower than $0.10$). In contrast, table 2 shows that the integration of GSEA with the three combination measures discover multiple significant gene sets in most of the cases. This difference implies that the discovered statistically significant gene sets include many moderate but coordinated differential gene combinations, even though the combinations are not significant by themselves as shown in table 1. This comparison demonstrates that traditional DGCS, similar to univariate gene analysis, has limited statistical power, and DGCS$\oplus$GSEA can increase that power.

\subsubsection{\textbf{GSEA-assisted DGCS has better result consistency than traditional DGCS}}

We further compare DGCS and DGCS$\oplus$GSEA by studying the consistency of their results on the first three data sets that are all from lung cancer studies, as done in \cite{subramanian2005gsa}. For DGCS, $M_1$ discovered $2$ genes on Michigan but nothing from Boston and Stanford; $M_2$ discovered $645$ combinations on Boston but only 1 and 2 from Michigan and Stanford, respectively, and there are no common ones between the 645, 1, and 2 gene combinations; $M_3$ discovered $10$ genes on Boston but only $1$ gene on Michigan and nothing from Stanford, and the 10 and 1 combinations do not overlap. The inconsistent results make the follow-up biological interpretation very difficult.

	In contrast, when the three DGCS measures are integrated with GSEA, several consistent themes can be observed: (i) Apoptosis related pathways (marked by $A$ in table 2): $M_1\oplus$GSEA discovered four gene sets on Boston, three of which are known to be closely related to cancer and specifically to apoptosis, i.e. \emph{nfkbpathway}, \emph{ST-Gaq-Pathway} and \emph{TNF-Pathway}. This apoptosis theme is shared by the gene sets discovered by $M_1$-GSEA from Michigan and Stanford, i.e. \emph{Monocyte-AD-Pathway}, \emph{hivnefPathway}, \emph{deathPathway} and \emph{caspasePathway}. These apoptosis related pathways are enriched with the lung cancer samples with good outcome, which makes sense biologically and also corresponds to the proliferation theme supported by the gene sets enriched with the samples with poor outcome as reported in \cite{subramanian2005gsa}. Several other examples of the result consistency, as indicated by other superscripts in Table 2, are in the technical report. This comparison demonstrates that traditional DGCS, like univariate gene analysis, has poor result consistency across the three lung cancer data sets, and DGCS$\oplus$GSEA can improve its consistency by integrating DGCS measures with GSEA.

\subsubsection{\textbf{GSEA-assisted DGCS with different DGCS measures complement each other}}

The number of significant gene sets discovered by the three versions of GSEA varies, i.e. $M_1\oplus$GSEA and $M_2\oplus$GSEA discovered a bit larger number of significant gene sets than $M_3\oplus$GSEA. However, $M_3\oplus$GSEA still discovered several gene sets that are not discovered by $M_1\oplus$GSEA or $M_2\oplus$GSEA, e.g. one gene set from the Michigan data set and three from the Stanford data set. This indicates that $M_1\oplus$GSEA, $M_2\oplus$GSEA and $M_3\oplus$GSEA have complementary perspectives, i.e. different combination measures capture different aspects of the difference between the phenotype classes (recall the two types of combinations in Figure \ref{fig:gb2005toy12}). This also demonstrates the proposed framework is general enough to integrate any type of DGCS with GSEA.

\subsection{Q2: Does DGCS-assisted GSEA improve GSEA?}

In this Section, we want to answer the question (Q2) of whether the integration of DGCS and GSEA can improve traditional GSEA. For this comparison, we consider the integration of DGCS and GSEA as a DGCS-assisted GSEA approach. We design three sets of comparisons. Firstly, we compare the traditional univariate-statistic based GSEA (denoted as $M_0\oplus$GSEA) with the integrative framework where one gene-combinations measure is used instead of $M_0$. Specifically, we compare the gene sets discovered by $M_0\oplus$GSEA with the gene sets discovered by $M_1\oplus$GSEA, $M_2\oplus$GSEA and $M_3\oplus$GSEA. Then, we compare $M_0\oplus$GSEA with the integrative framework where one gene-combinations measure is used in addition to $M_0$, i.e. $M_{01}\oplus$GSEA, $M_{02}\oplus$GSEA and $M_{03}\oplus$GSEA. Furthermore, we also study the integration of multiple gene-combinations measure in addition to $M_0$, e.g. $M_{0123}\oplus$GSEA.

\begin{figure}[t]%
\centering
\includegraphics[width=.99\textwidth]{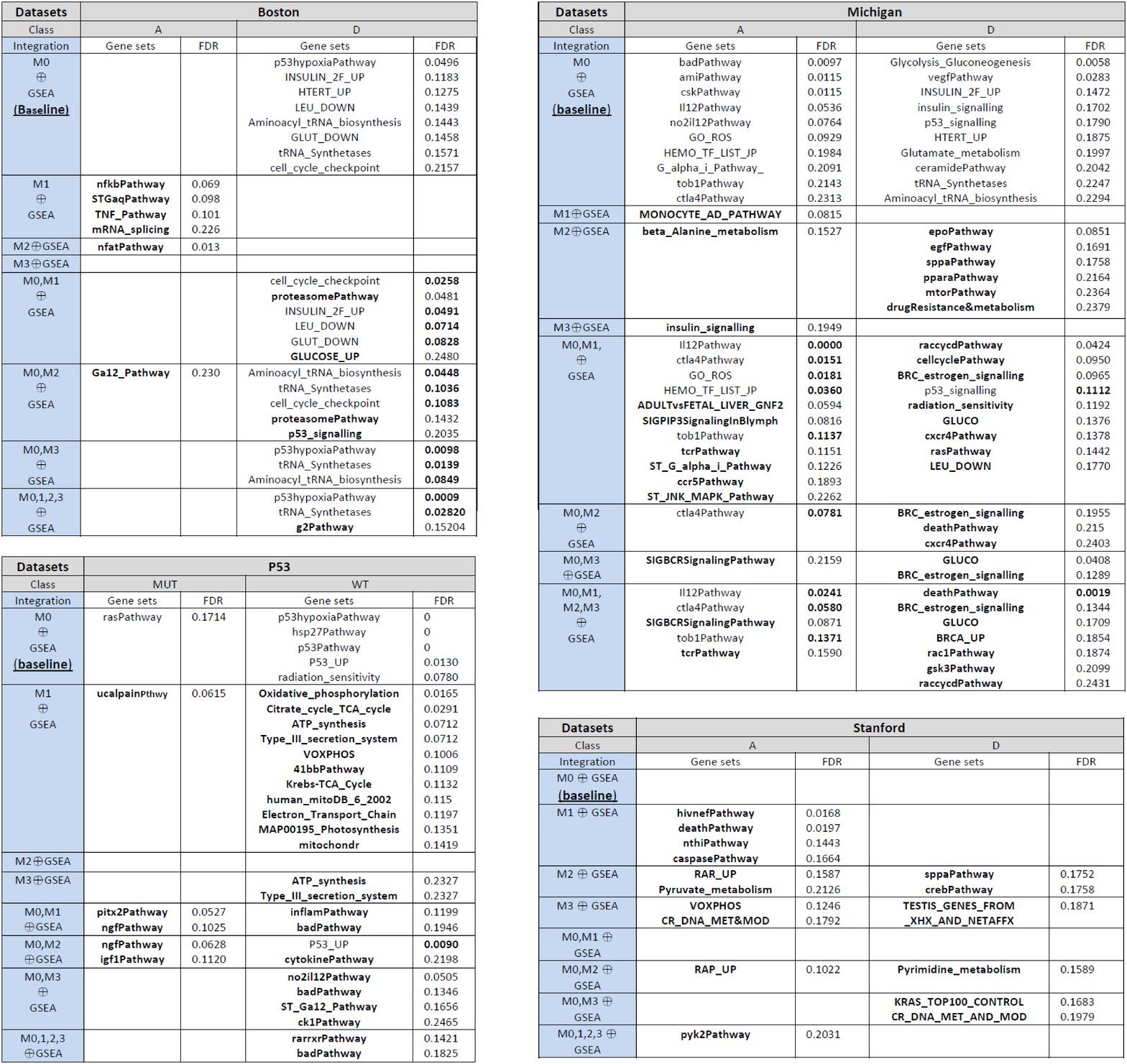}
\caption{\textbf{Common captions for the four tables}: Statistically significant gene sets discovered by different (combinations of) measures from each of the four data sets. The first row of each table shows the name of the data set, and the second row indicates the two phenotype classes in the data set that a gene set can be enriched with. The first column indicates the measures used in the integrative framework. For each data set and each (combination of) measure(s), we list the names of the statistically significant gene sets and the corresponding FDRs for both the classes. The traditional univariate-statistic based GSEA (M0$\oplus$GSEA) is considered as the baseline. For the other rows, we only list a gene set if it is only discovered by the integrative approach (with bolded name), or it has a non-trivially decreased FDR when it is discovered by the integrative approach (with bolded FDR). Please refer to the Appendix section for the complete tables.} %
\label{fig:bigtable12312009half}
\end{figure}

Figure \ref{fig:bigtable12312009half} displays the statistically significant gene sets discovered with different (combinations of) measures respectively from the four datasets. An FDR threshold of $0.25$ is used as in \cite{subramanian2005gsa} for comparison purpose. The results presented in \cite{subramanian2005gsa} are exactly reproduced, i.e. the gene sets listed in the rows corresponding to $M_0\oplus$GSEA. In each of these four figures, we consider the traditional univariate-statistic based GSEA ($M_{0}\oplus$GSEA) as the baseline, and compare it with the rows corresponding to $M_{1}\oplus$GSEA, $M_{2}\oplus$GSEA, $M_{3}\oplus$GSEA, $M_{01}\oplus$GSEA, $M_{02}\oplus$GSEA, $M_{03}\oplus$GSEA and $M_{0,1,2,3}\oplus$GSEA. From these comparisons, the following observations can be made. %

\subsubsection{\textbf{DGCS-assisted GSEA discovers additional significant gene sets}}

First, we compare the rows corresponding to $M_{1}\oplus$GSEA, $M_{2}\oplus$GSEA, $M_{3}\oplus$GSEA with the rows corresponding to $M_{0}\oplus$GSEA. We bolded the additional gene sets that are only discovered by $M_{1}\oplus$GSEA, $M_{2}\oplus$GSEA, $M_{3}\oplus$GSEA. For example, with $M_0\oplus$GSEA, no statistically significant gene sets have
been enriched with class A in the Boston data set. In contrast, $M_1\oplus$GSEA discovered $4$ gene sets, three out of
which (discussed in $Q1$) are related to apoptosis which is consistent with the results on Michigan and Stanford. On the Michigan dataset, $M_2\oplus$GSEA discovered a gene set \emph{beta-Alanine-metabolism} that is not discovered by
$M_0\oplus$GSEA. This gene set is related to the responses of hypoxia, which is consistent with the results on Boston and Stanford. It is worth noting that, although most studies did not report statistically significant gene sets on the Stanford dataset due to the very small sample size, $M_{1}\oplus$GSEA, $M_{2}\oplus$GSEA, $M_{3}\oplus$GSEA respectively discovered 4, 4 and 3 significant gene sets. These additional gene sets were discovered because the three DGCS measures capture different types of the differentiation between the two phenotype classes, compared to the traditional univariate differential expression-based GSEA.

Second, we compare the rows corresponding to $M_{01}\oplus$GSEA, $M_{02}\oplus$GSEA, $M_{03}\oplus$GSEA with the rows corresponding to $M_{0}\oplus$GSEA. We bolded the additional gene sets that are only discovered by the integrative approach. For example, on the Boston data set, $M_0$ based GSEA discovered 8 gene sets. In addition, $M_{01}\oplus$GSEA discovered the \emph{proteasomePathway} gene set, and $M_{02}\oplus$GSEA discovered the \emph{p53-signaling} gene set. Both ubiquitin-proteasome pathway and p53-signaling pathway are well-known cancer-related pathways that are also specifically related to lung cancer \cite{li2009phaseproteasome,chenette2009signallingLung}. (Additional examples are in the technical report.) The gene sets that are discovered by DGCS-assisted GSEA but not by $M_{0}$-GSEA illustrate the benefits of using DGCS to assist GSEA.

Next, we also observed that integrating multiple DGCS measures can further discover statistically significant gene sets. For illustration purpose, we compare the rows corresponding to $M_{01}\oplus$GSEA, $M_{02}\oplus$GSEA, $M_{03}\oplus$GSEA with the rows corresponding to $M_{0123}\oplus$GSEA. $M_{0123}\oplus$GSEA discovers the \emph{g2Pathway} gene set and the \emph{gsk3Pathway} gene set, respectively from the Boston and the Michigan dataset. Neither of these two pathways are discovered by $M_{0}\oplus$GSEA, $M_{01}\oplus$GSEA, $M_{02}\oplus$GSEA and $M_{03}\oplus$GSEA. The curated gene set \emph{g2Pathway} contains the genes related to the G2/M transition, which is shown to be regulated by p53 \cite{taylor2001regulationG2M}, a well-known cancer-related gene. The curated gene set \emph{gsk3Pathway} is the signaling pathway of GSK-3-$\beta$, which has been shown to be related to different types of cancer\cite{cao2006glycogen,liao2004glycogen}. These two cancer-related pathways are discovered by $M_{0123}\oplus$GSEA but not by $M_{0}\oplus$GSEA, $M_{01}\oplus$GSEA, $M_{02}\oplus$GSEA and $M_{03}\oplus$GSEA. This indicates that different members of these two pathways are differential between the two phenotype groups in different manners, i.e. the differentiation of some genes is captured by $M_0$, some by $M_1$, some by $M_2$ and some by $M_3$. These two pathways can be discovered to be statistically significant only when these measures are used together in the integrative framework. This demonstrates the benefits of the proposed framework for integrating multiple DGCS measures with a univariate measure. %

It is worth noting that, the gene sets discovered by the integrative framework with multiple measures are not necessarily a superset of those discovered by integrating each individual measure with GSEA since, when different DGCS measures are integrated with GSEA, the null-hypotheses tested in the GSEA framework are correspondingly different. The highlight of the integrative framework is that, additional gene sets can be discovered when different DGCS measures are used to assist the traditional univariate statistic-based GSEA. In practice, these different versions of GSEA should be used collectively.

\subsubsection{\textbf{DGCS-assisted GSEA discovers gene sets with lower FDRs}}:

Even when a gene set is discovered both before and after a DGCS measure is integrated into the framework, we can observe several interesting cases where the FDR of a gene set becomes much lower after the integration. We bolded the FDRs that significantly decreased when they are discovered by the integrative approach. For example, $M_{0123}\oplus$-GSEA, in which $M_0$, $M_1$, $M_2$ and $M_3$ are integrated together, discovers \emph{p53hypoxialPathway} with an much lower FDR of $0.00095$, two-order lower than $M_0$-GSEA. This example indicates that several members of \emph{p53hypoxialPathway} have weak individual differentiation measured by $M_0$, but have more significant differentiation when they are measured by $M_3$. This and other similar examples demonstrates the benefits of the proposed framework for integrating multiple DGCS measures.

\subsubsection{\textbf{DGCS-assisted GSEA further improve the consistency across the three lung cancer data sets}}

As presented in \cite{subramanian2005gsa}, $M_0\oplus$GSEA discovered $8$ and $11$ gene sets respectively from the Boston and Michigan data sets, and 5 of the 8 in Boston and 6 of the 11 in Michigan are common. The three unmatched gene sets that are discovered in Boston but not in Michigan are \emph{GLUT-DOWN}, \emph{LEU-DOWN} and \emph{CellCycleCheckpoint}. Interestingly, the latter two are discovered from both the Boston and the Michigan data sets by $M_{01}\oplus$GSEA\footnote{The \emph{CellCyclePathway} discovered on Michigan and the \emph{cell-cycle-checkpoint} discovered on Boston are both cell-cycle related gene sets}. Such observations suggest that DGCS-assisted GSEA also provides new insights to the consistency between different data sets.

\subsubsection{\textbf{Additional issues of multiple hypothesis testing}}

Because different combinations of measures are used in the integrative framework, additional issues of multiple hypothesis testing arise, even though multiple hypothesis testing has been addressed for each measure via the phenotype permutation test procedure in the GSEA framework proposed in \cite{subramanian2005gsa}. To investigate this, we designed experiments with 4 of the $15$$(=2^4-1)$ possibilities of integrations, i.e. $M_{01}\oplus$GSEA, $M_{02}\oplus$GSEA, $M_{03}\oplus$GSEA and $M_{0123}\oplus$GSEA. Even using a collective (meta-level) multiple hypothesis correction, many discovered gene sets would still be significant. For examples, $M_{0123}\oplus$GSEA discovers \emph{p53hypoxialPathway} from the Boston data set with a low FDR of $0.00095$, and $M_{0123}\oplus$GSEA discovers \emph{deathPathway} from the Michigan data set with a lower FDR of $0.00197$. We also did additional permutation tests, in which we generate random gene sets with the same sizes as the sets in MSigDB $C_2$, and do the same set of experiments as shown in Figure \ref{fig:bigtable12312009half}. The FDR values of the random gene sets computed in the integrative framework are mostly insignificant (higher than $0.25$).%


%
\section{Discussion}
\label{sec:discssion}

In this paper we motivated the integration of differential gene-combination search and gene set enrichment analysis for bi-directional benefits on both them. We proposed a general integrative framework that can handle gene-combinations of different sizes ($k \geq 2$) and different gene-combination measures in addition to an univariate statistic used in traditional GSEA. The experimental results demonstrated that, on one hand, GSEA-assisted DGCS has better statistical power and result consistency than traditional DGCS. On the other hand,  DGCS-assisted GSEA can discover additional statistically significant gene sets that are ignored by traditional GSEA and further improve the result consistency of the traditional GSEA.

The proposed framework can be extended in several ways. Different variations of GSEA will be considered. Along these lines, we note that the proposed integrative framework is general enough to integrate most existing variations of GSEA approaches summarized in \cite{ackermann2009gmf261} with minimal amount of modification. Also, it should be possible to integrate DGCS and gene-subnetwork discovery. Both GSEA and gene-subnetwork discovery \cite{ideker2002discovering,subnetwork2007} can discover collections of genes, either known gene sets \cite{subramanian2005gsa} or subnetworks in a molecular network (e.g. protein interaction network), that show moderate but coordinated differentiation. In this paper, we integrate DGCS and GSEA as an illustration of the general framework for integrating scores from different gene-combination measures and gene-combinations of different sizes, in addition to the traditional univariate statistic, but the same framework also applies to the integration of DGCS and gene subnetwork discovery. Another direction is the use of this framework for the analysis of (GWAS) SNP data, by following the methodology proposed in recently work on pathway/network based analysis of GWAS datasets \cite{wang2007pathway,baranzini2009pathway}. Finally, it may be possible to use constraints on gene-combinations to improve our framework. In procedure $A$, for each gene-combination measure and an integer $k$, the score of a gene is assigned from all the ${N-1 \choose k-1}$ possible gene-combinations. A further extension of procedure $A$ is to only consider the gene combinations, in which the $k$ genes appear in a common gene set, e.g. a pathway. Such gene-set-based constraints may better control false positive gene combinations and improve the statistical power of the whole integrative framework.

\section{Appendix}

\subsection{Illustration of procedure $A$}

\begin{figure}[t]%
\centering
\includegraphics[width=.85\textwidth]{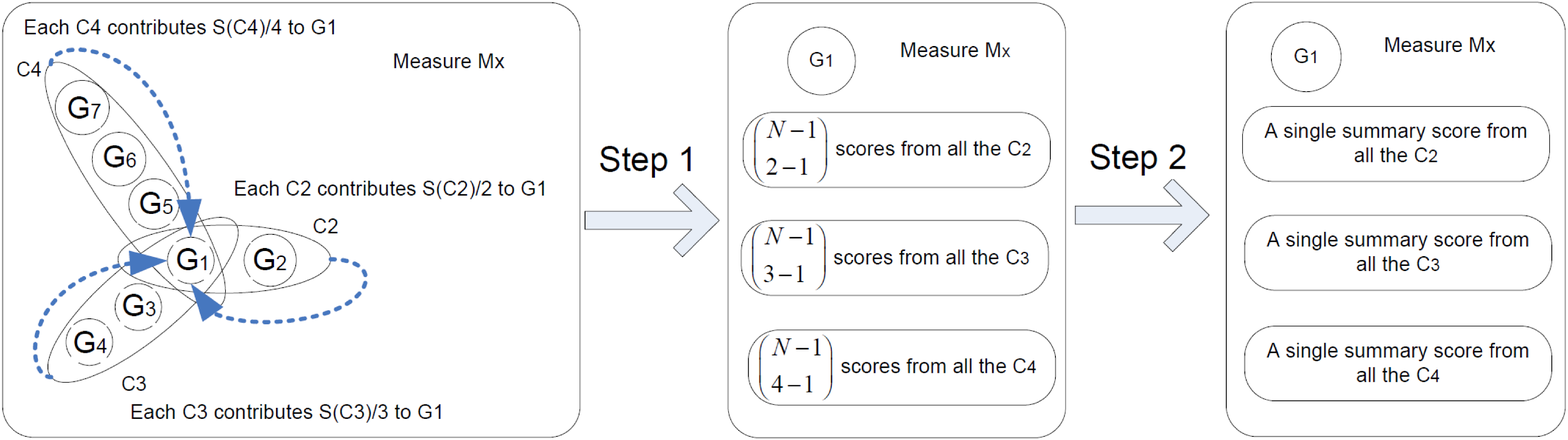}
\caption{Illustration of procedure $A$ (combination-to-gene score assignment) for $G_1$. The three ellipses represent the three gene combinations that $G_1$ participates to with respect to measure $M_1$.} %
\label{fig:comb2gene}
\end{figure}

The illustration of procedure $A$ is in figure \ref{fig:comb2gene}.

\subsection{Complete Gene-set Table}

Due to the space limit, Table 2 and Figure \ref{fig:bigtable12312009half} are both summarized from the four complete tables that are available at http://vk.cs.umn.edu/ICG/. Specifically, Table 2 is a high-level summary of the number of gene sets discovered and the biological processes associated with each of the gene sets. In figure \ref{fig:bigtable12312009half}, we listed the complete results for $M_0\oplus$GSEA (the baseline), while for the other rows, we only list a gene set if it is only discovered by the integrative approach (with bolded name), or it has a non-trivially decreased FDR when it is discovered by the integrative approach (with bolded FDR).

%
%
%
%
%
%
%
%
%
%
%
%
%
%
%
%
%
%
%
%
%
%
%
%
%
%
%
%
%
%
%

%

%
%
%
%
%
%
%
%
%
%


\end{document}